\documentclass[conference]{IEEEtran}
\IEEEoverridecommandlockouts
\usepackage{kotex}
\usepackage{booktabs}
\usepackage{subcaption}
\usepackage{cite}
\usepackage{amsmath,amssymb,amsfonts}
\usepackage{algorithmic}
\usepackage{graphicx}
\usepackage{textcomp}
\usepackage{xcolor}
\def\BibTeX{{\rm B\kern-.05em{\sc i\kern-.025em b}\kern-.08em
    T\kern-.1667em\lower.7ex\hbox{E}\kern-.125emX}}
\begin{document}

\title{Wafer-Level Etch Spatial Profiling for Process Monitoring from Time-Series with Time-LLM
% Wafer-Level Etch Profiling for In-Situ Process Monitoring Using a Time-LLM–Based Approach
%\title{Wafer-Level Etch Profiling for In-Situ Process Monitoring Using Large Language Models
%
%\title{Time-LLM–Based Wafer-Level Etch Profiling \\ for In-Situ Process Monitoring
%\title{Time-LLM-Based Prediction of Wafer-Level Etch Quality from In-Situ Process Surveillance \\
%\title{Time-LLM-Based Prediction of Wafer-Level Etch Quality from In-Situ Process Time-Series Signals \\
%\textcolor{red}{surveillance 느낌 추가}

%\title{Time-LLM-based Wafer-Level Etch Spatial Profile Prediction from In-situ Process Time Series\\
% {\footnotesize \textsuperscript{*}Note: Sub-titles are not captured for https://ieeexplore.ieee.org  and should not be used}

\author{

\IEEEauthorblockN{
Hyunwoo Kim\IEEEauthorrefmark{1},
Munyoung Lee\IEEEauthorrefmark{1},
Seung Hyub Jeon,
Kyu Sung Lee
\thanks{\IEEEauthorrefmark{1} These authors contributed equally to this work.}
\thanks{This work was supported by the Technology Innovation Program (Development and Practice of an On-device AI Functionality and Performance Testing Framework based on NPU, RS-2025-02307650) funded by the Ministry of Trade Industry and Resources (MOTIR, Korea) and ETRI grant funded by the Korea government (26ZT1100, 26YT1100).}
}

\IEEEauthorblockA{
%\textit{Regional ICT Convergence Research Section} \\
%\textit{Electronics and Telecommunications Research Institute}\\
Electronics and Telecommunications Research Institute\\
Daejeon, Republic of Korea \\
\{kim.hw, munyounglee, shjeon00, kyusung.lee\}@etri.re.kr}
}

}
% \and
% \IEEEauthorblockN{
% Munyoung Lee\textsuperscript{*}
% \thanks{* These authors contributed equally to this work.}}
% \IEEEauthorblockA{
% \textit{Regional ICT Convergence Research Section} \\
% \textit{Electronics and Telecommunications Research Institute}\\
% Daejeon, Republic of Korea \\
% munyounglee@etri.re.kr}
% \and

% \IEEEauthorblockN{
% 3\textsuperscript{rd} Given Name Surname}
% \IEEEauthorblockA{
% \textit{dept. name of organization (of Aff.)} \\

% \and
% \IEEEauthorblockN{
% 4\textsuperscript{th} Given Name Surname}
% \IEEEauthorblockA{
% \textit{dept. name of organization (of Aff.)} \\

\maketitle

\begin{abstract}
Understanding wafer-level spatial variations from in-situ process signals is essential for advanced plasma etching process monitoring.
While most data-driven approaches focus on scalar indicators such as average etch rate, actual process quality is determined by complex two-dimensional spatial distributions across the wafer.
This paper presents a spatial regression model that predicts wafer-level etch depth distributions directly from multichannel in-situ process time series.
We propose a Time-LLM–based spatial regression model that extends LLM reprogramming from conventional time-series forecasting to wafer-level spatial estimation by redesigning the input embedding and output projection.
Using the BOSCH plasma-etching dataset, we demonstrate stable performance under data-limited conditions, supporting the feasibility of LLM-based reprogramming for wafer-level spatial monitoring.

\end{abstract}


\section{Introduction}

%최근 딥러닝과 대규모 언어 모델(LLM)의 발전으로, 반도체 제조 공정의 실시간 감시와 모니터링을 위한 데이터 기반 연구가 활발히 이루어지고 있다.
%특히 Virtual Metrology (VM), Fault Detection and Classification (FDC), Advanced Process Control (APC), Prognostics and Health Management (PHM) 등은 장비 및 공정 상태를 실시간으로 추정하여 수율 저하나 공정 이상을 조기에 탐지하는 핵심 기술로 자리 잡고 있다~\cite{MAITRA2024VM-survey, Kim2022FDC, Jalali2019PHM}.
%분량 이슈시 FDC, PHM 레퍼는 삭제 ~\cite{MAITRA2024VM-survey}.
%\textcolor{red}{(VM외에도 FDC, PHM 서베이 논문 추가가 필요할까요?)-그냥 대표적인 논문 여기에 레퍼런스만 다는 수준이면 될거 같아요.}
% 이들 접근법의 공통된 목표는 공정 상태를 사전에 파악하고 변동을 조기에 감지함으로써 제조 안정성과 수율을 향상시키는 데 있다.

Recent advances in deep learning and large language models (LLMs) have enabled data-driven real-time monitoring in semiconductor manufacturing. Specifically, methodologies such as virtual metrology (VM), fault detection and classification (FDC), and prognostics and health management (PHM) play a pivotal role in detecting yield excursions and process anomalies by leveraging in-situ process data~\cite{MAITRA2024VM-survey, Kim2022FDC, Jalali2019PHM}.
Deposition and etching are critical processes that determine device geometry, and the in-situ time-series data collected during these steps serve as key indicators of process state.
However, conventional in-situ signals primarily capture temporal dynamics and do not directly reflect intra-wafer spatial variations.
As a result, predicting wafer-level spatial profiles from in-situ signals has become essential, as these profiles directly characterize uniformity, yield, and process drift.

% Deposition and etching are critical processes that determine device geometry, and the in-situ time-series data collected during these steps serve as key indicators of process state. However, conventional time-series signals mainly capture temporal dynamics and fail to reflect spatial variations within a wafer. As a result, predicting wafer-level spatial profiles from in-situ signals has become essential, as these profiles directly characterize uniformity, yield, and process drift, linking monitoring data to wafer-level etch distributions.

%반도체 제조에서 증착과 식각 공정은 소자 형상을 직접적으로 결정하는 핵심 단계이며, 이 과정에서 수집되는 in-situ 공정 시계열 데이터는 공정 상태를 반영하는 주요 정보원이다.
%그러나 이러한 시계열 신호는 공정의 시간적 거동을 중심으로 나타낼 뿐, wafer 내부의 위치별 편차나 공간적 비균일성을 직접적으로 설명하는 데에는 한계가 있다.
%이에 따라, in-situ 시계열로부터 wafer-level 공간 프로파일을 예측하려는 접근이 주목받고 있다.
%Wafer-level 공간 프로파일은 균일도, 수율, 드리프트를 직접적으로 반영하므로, 공정 감시 정보를 실제 wafer 수준의 식각 분포와 정량적으로 연결하는 핵심 지표로 활용될 수 있다.

Despite this importance, wafer-level spatial profile prediction remains limited. Prior studies largely focus on scalar indicators such as average etch rate or on coarse spatial descriptors using statistical or deep learning models~\cite{YAO-2025-Profiling, pamarty-2024-Profiling, Kota-2025-Profiling}, while end-to-end regression of wafer-level two-dimensional spatial fields from in-situ time-series data has been rarely explored. In addition, many existing approaches rely on proprietary and small-scale datasets, which hampers systematic evaluation of generalization across diverse process conditions.

Recently, LLM-based time-series models pretrained on large-scale datasets have emerged as an alternative.
Time-LLM~\cite{jin2023time} reprograms time-series inputs into the LLM token space via patch-based input transformations and conditions the model using domain knowledge and statistical priors through a Prompt-as-Prefix mechanism.
By freezing the LLM backbone and optimizing only lightweight reprogramming modules, the framework achieves stable learning even in data-scarce regimes.
However, Time-LLM is fundamentally designed for forecasting within the same numerical domain and cannot be directly applied to spatial regression tasks that map temporal in-situ signals to wafer-level two-dimensional profiles.

% To address these limitations, large-scale pre-trained language models have recently been adapted for time-series analysis.
% Time-LLM~\cite{jin2023time} is a framework that reprograms time-series data into the LLM embedding space by applying patch-based input transformation and a Prompt-as-Prefix mechanism, enabling the incorporation of domain knowledge and statistical features.
% By training only a lightweight reprogramming module while keeping the LLM parameters fixed, Time-LLM achieves high sample efficiency and stable learning, even in data-limited scenarios.
% However, as Time-LLM was originally designed for temporal forecasting tasks, it is not readily applicable to spatial regression problems that require mapping in-situ time-series signals to continuous wafer-level two-dimensional spatial fields.

%이러한 한계를 극복하기 위한 대안으로, 대규모 데이터로 사전 학습된 LLM 기반 시계열 모델이 주목받고 있다.
%Time-LLM~\cite{jin2023time}은 시계열 데이터를 LLM embedding 공간으로 재정렬하여 분석하는 프레임워크로, patch 기반 입력 변환과 Prompt-as-Prefix를 통해 도메인 정보와 통계적 특성을 주입한다.
%LLM을 고정한 상태에서 경량 reprogramming 모듈만 학습하므로, 데이터가 제한된 환경에서도 안정적인 학습이 가능하다.
%그러나 Time-LLM은 동일한 수치 도메인에서 미래 값을 예측하는 시계열 forecasting 문제를 대상으로 설계되었으며, in-situ 시계열에서 wafer-level 2-D 공간 도메인으로의 매핑을 요구하는 공간 회귀 문제에는 그대로 적용하기 어렵다.

In this paper, we present a Time-LLM–based spatial regression framework for inferring wafer-level etch depth distributions directly from in-situ process parameter and optical emission spectroscopy (OES) time series.
By systematically redesigning the input embedding and output projection layers, the proposed approach extends the Time-LLM paradigm beyond conventional temporal forecasting to support direct time-series-to-spatial regression.
% To extend the original Time-LLM architecture beyond time-series forecasting, we redesign the input embedding and output projection layers to enable direct mapping from temporal sequences to continuous two-dimensional spatial fields.
% The proposed model is validated using a real-world BOSCH plasma-etching dataset. Experimental results demonstrate that the model achieves accurate wafer-level spatial predictions and robust generalization under process drift, even in data-limited conditions. 
Experimental evaluation on the BOSCH plasma-etching dataset demonstrates accurate reconstruction of wafer-level spatial profiles and robust generalization under process drift, even in data-limited regimes.
% The model is validated using the BOSCH plasma-etching dataset, demonstrating accurate spatial predictions and robust generalization under process drift, even in data-limited conditions.
To the best of our knowledge, this work constitutes the first empirical study to perform wafer-level spatial profile prediction on the real-world BOSCH plasma-etching dataset, underscoring the potential of LLM-based reprogramming for advanced semiconductor process monitoring.
%To the best of our knowledge, this is the first study to demonstrate wafer-level spatial profile prediction on a real-world BOSCH plasma-etching dataset using an LLM-based reprogramming framework, suggesting that LLM-based reprogramming offers a scalable and effective approach for semiconductor process monitoring.

%본 논문에서는 in-situ 공정 시계열 데이터(process parameters 및 optical emission spectroscopy)로부터 wafer-level etch spatial profile을 예측하는 Time-LLM 기반 공간 회귀 모델을 제안한다.
%기존 시계열 forecasting을 목적으로 설계된 Time-LLM 구조를 확장하기 위해, 입력 embedding과 output projection을 재구성하여 시계열-to-공간 회귀 문제에 적합한 모델 구조를 설계하였다.
%제안한 모델은 실제 공정에서 수집된 BOSCH plasma-etching 데이터셋을 활용하여 검증되었다.
%실험 결과, 제안한 모델은 제한된 데이터 환경에서도 wafer-level 공간 프로파일 예측 정확도와 공정 드리프트에 대한 견고성을 동시에 확보함을 확인하였다.
%이러한 결과는 LLM 기반 reprogramming 접근이 wafer-level 공간 정보를 활용한 반도체 공정 모니터링에 효과적이고 확장 가능한 방법임을 실증적으로 보여준다.
%아는 한도 내에서, 본 연구는 BOSCH plasma-etching 데이터셋을 활용하여 wafer-level spatial profile 예측을 수행한 최초의 연구이다.

% \textcolor{red}{실제 공정을 통해 수집한 BOSCH plasma-etching 데이터셋을 활용한 초기 연구 사례로, 실험을 통해}
% 제한된 데이터 환경에서도 LLM 기반 reprogramming 접근이 반도체 공정 감시 문제에 효과적으로 적용될 수 있음을 확인하였다.
% \textcolor{red}{(contribution 및 result 추가) 
% 제안하는 방식은 제한된 데이터 환경에서도 웨이퍼 수준 공간 프로파일 예측 정확도 측면에서 우수한 예측 성능과 드리프트에 대한 견고성을 보임을 확인하였다. 이러한 결과는 LLM 기반 재프로그래밍 접근 방법이 반도체 공정 감시 응용에 효과적으로 활용될 수 있음을 보여준다.}

\section{Related Work}

% 대부분 기업 내부 비공개 데이터로, 특정 장비·node에 국한된 소규모 데이터셋을 사용.
% public benchmark가 거의 없어, 모델 간 성능 비교나 일반화 성능 평가가 어려움.
% “plasma 기반 공정에서 AI를 사용한 사례는 다수 존재하지만, wafer 2‑D spatial profile을 연속적으로 예측하는 문제는 아직 공백이 있음
% 실제 plasma 기반 공정(etch, deposition, CMP 등)에 대해 공정 조건을 입력으로 wafer 전체 2‑D spatial profile을 end‑to‑end로 예측하는 ML/DL/LLM 기반 연구 잘 없음-> 기존에는 스칼라 지표나 wafer map 패턴 분류에 한정되었던 AI 응용을, 연속 2‑D 프로파일 예측까지 확장한 최초 시도 주장?
% Prior works on wafer spatial modeling (e.g., GPR with sampling or AutoML ensembles ) excel in interpolation from limited test data but fail to predict full 2D profiles from plasma process recipes due to data scarcity and limited reasoning. Wafer map classification via CNNs addresses patterns but not continuous physics-based profiles. Thus, pre-trained LLMs, with strong generalization from vast data, are essential for overcoming these limitations.

\subsection{Virtual Metrology for Process Monitoring}
%반도체 제조의 미세화에 따라 실시간 공정 감시를 위한 VM 연구가 관심받고 있음.
%초기 VM 연구는 장비 데이터와 OES(Optical Emission Spectroscopy) 센서를 결합하여 박막 두께나 식각율의 평균값을 예측 
%최근에는 플라즈마 정보(PI) 변수를 활용하여 식각 깊이나 CD 예측까지 확장 
% 그러나 이러한 접근은 단일 지점의 공정 결과 값 예측에 머물러 있어, 웨이퍼  불균일성 모니터링하기에는 부족함 
% 공정 변화에 따른 데이터 분포 차이(Domain shift)를 극복하기 위해 도메인 적응(Domain Adaptation) 기법이 제안되었으나 2D 공간 변동 포착 한계 

With the increasing complexity of semiconductor manufacturing, virtual metrology has become an important technique for real-time process monitoring in yield management~\cite{MAITRA2024VM-survey}.
Prior studies utilized in-situ sensor data such as OES to predict scalar process outputs, including mean etch rate and film thickness~\cite{CHOI2021VM}.
% VM studies have utilized in-situ sensor data, such as optical emission spectroscopy (OES), to predict scalar process outputs including mean etch rate and film thickness~\cite{CHOI2021VM}.
Other studies extended these approaches to estimate etch profile parameters, such as etch depth and bowing critical dimension, by extracting plasma information variables~\cite{MAITRA2024VM-survey, Kwon2021VM}.
However, most VM approaches remain focused on point-wise predictions, limiting their ability to capture dense wafer-level spatial non-uniformity.

\subsection{Data-Driven Spatial Etch Profiling}
% 웨이퍼 레벨의 2D 공간 프로파일을 모델링하기 위해 통계적 기법부터 최신 딥러닝까지 다양한 시도 
% 과거에는 통계적 방식이 사용되었으나 최근에는 Cascade RNN(CRNN)을 통한 고속 프로파일 예측 [13]이나, Autoencoder와 RBF(Radial Basis Function) 보간법을 결합한 하이브리드 모델 [14],  물리 기반 합성 데이터로 학습된 신경망 서로게이트 모델 [15] 등이 제안 
% 하지만 이러한 모델들은 대부분 특정 챔버의 제한된 데이터에 의존하여 학습되므로 일반화 성능이 낮으며, 데이터가 부족한 환경에서 성능이 급격히 저하되는 오버피팅(Overfitting) 문제 
% 특히 연속적인 2D 공간 필드(Field) 예측은 단순한 결함 패턴 분류 [19, 21]보다 훨씬 높은 복잡도 요구

The modeling of wafer-level two-dimensional spatial profiles has evolved from classical statistical methods to modern data-driven deep learning approaches.
Representative examples include Cascade RNNs (CRNNs)~\cite{YAO-2025-Profiling}, hybrid Autoencoder–RBF models~\cite{pamarty-2024-Profiling}, and neural surrogate models trained on physics-based synthetic data~\cite{Kota-2025-Profiling}. 
Some studies further divided wafers into regions and predicted regional thickness uniformity from plasma signals~\cite{lee-2024-edge_region}.
% In addition, wafers were segmented into multiple spatial regions (center, edge, and ring), and independent models were trained on plasma information (PI) signals to predict regional thickness uniformity~\cite{lee-2024-edge_region}.
%In addition, wafers have been divided into multiple spatial regions (e.g., center, edge, and ring), and independent models were applied to plasma information (PI) sensor signals to predict etch rate or regional thickness uniformity~\cite{lee-2024-edge_region}.
Nevertheless, dense multi-point spatial regression remains challenging due to limited labeled data and increased model complexity.

\subsection{Time-Series-to-Spatial Regression via LLMs}

Recent time-series research has demonstrated that frozen LLMs can serve as effective temporal feature extractors and enable few-shot forecasting~\cite{jin2023time, zhou2023onefitsall, Chang2025LLM4TS, gruver2023llmtime, ansari2024chronos}.
% Recent time-series research has increasingly leveraged pre-trained LLMs for representation learning.
% Prior works demonstrated that frozen LLMs can serve as powerful temporal feature extractors and enable effective few-shot forecasting~\cite{jin2023time, zhou2023onefitsall, Chang2025LLM4TS, gruver2023llmtime, ansari2024chronos}.
However, these approaches primarily address temporal forecasting within the same numerical domain, whereas wafer etch profile prediction requires mapping time-series signals to two-dimensional spatial outputs.
In constrast, this work adapts LLM-based reprogramming to this cross-domain regression problem by restructuring the output layer for wafer-level spatial prediction.

\section{Background}

\subsection{Baseline model: Time-LLM}   \label{sec:timellm}

\begin{figure}[t]
  \centering
  \includegraphics[width=\linewidth]{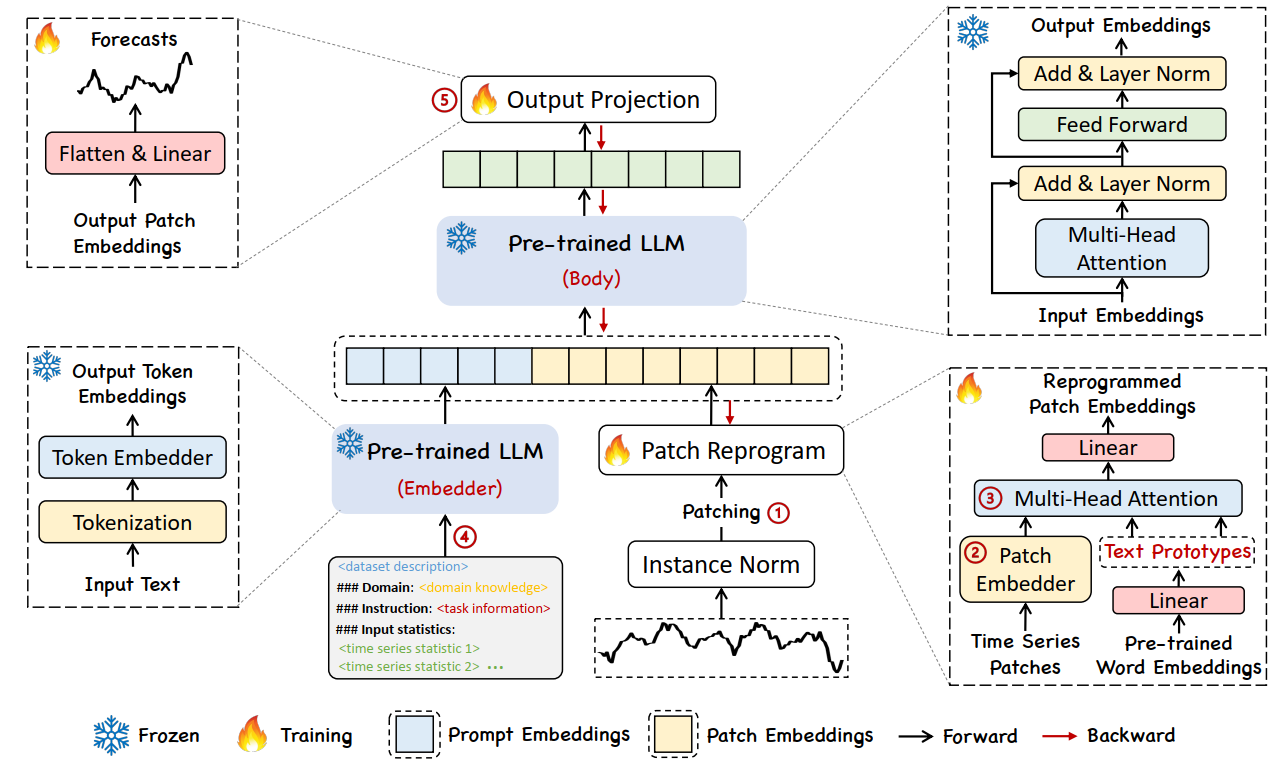}
%  \caption{Time-LLM 구조 그림}
  \caption{Overall architecture of the Time-LLM framework, reproduced from~\cite{jin2023time}.}
  \label{fig:timellm_arch}
\end{figure}

% Time-LLM은 사전학습된 대규모 언어 모델(LLM)을 고정된 backbone으로 활용하여 시계열 분석을 수행하는 forecasting 프레임워크이다.
% 구조는 Fig.~\ref{fig:timellm_arch}와 같으며, 입력 시계열은 patch 단위로 분할된 후 patch embedding과 cross-attention 기반 reprogramming 과정을 거쳐 LLM의 embedding 공간으로 사상되며, Prompt-as-Prefix(PaP)를 통해 데이터 특성과 예측 과제가 자연어 형태로 주입된다.
% LLM은 reprogrammed 시계열 토큰과 PaP를 함께 처리하여 시간적 패턴에 대한 고차원 표현을 생성하고, output projection 단계에서는 해당 표현을 다시 수치 도메인으로 변환하여 예측 출력을 생성한다.
% 이 과정에서 LLM의 파라미터는 고정된 상태로 유지되며, reprogramming 및 projection과 같은 경량 모듈만을 학습함으로써 데이터가 제한된 환경에서도 안정적인 학습이 가능하다.

Time-LLM~\cite{jin2023time} is a forecasting framework that employs an LLM as a fixed backbone for time-series analysis.
As shown in Fig.~\ref{fig:timellm_arch}, the input time series is partitioned into patches and reprogrammed into the LLM embedding space via patch embedding and cross-attention, while data characteristics and task instructions are injected through a Prompt-as-Prefix (PaP) mechanism.
The LLM processes the reprogrammed tokens together with the PaP to capture temporal patterns, and the resulting representations are mapped back to the numerical domain by an output projection layer to generate predictions.
By freezing the LLM parameters and training only lightweight reprogramming and projection modules, Time-LLM enables stable learning in data-limited settings.

\subsection{Target Domain: Silicon Etching Process}

% 플라즈마 기반 식각 공정은 플라즈마 내 이온과 라디칼의 물리·화학적 반응을 이용해 방향성 있는 식각을 수행하는 반도체 제조의 핵심 공정으로, 공정 결과는 플라즈마 챔버의 동작 조건에 민감하게 의존한다.
% 공정 거동은 gas flow, RF power, chamber pressure, 온도와 같이 플라즈마 챔버 상태를 직접 제어하는 process parameters에 의해 결정되며, 이들은 공정 제어 및 모니터링을 위해 in-situ 시계열 데이터로 수집된다.
% 한편, 플라즈마 상태는 공정 변수만으로 완전히 설명되기 어렵기 때문에, optical emission spectroscopy(OES)이 함께 활용된다.
% OES는 플라즈마 방출 스펙트럼을 통해 활성 반응종의 거동과 플라즈마 상태 변화를 실시간으로 관측할 수 있는 수단을 제공한다.
% 결과적으로, process parameter와 OES로 구성된 다채널 in-situ 시계열은 공정 상태를 반영하는 핵심 정보원이며, 이로부터 wafer-level 식각 분포와 공간적 비균일성을 예측하는 것은 공정 모니터링과 이상 감지를 위한 중요한 과제이다.

Plasma-based etching is a core semiconductor manufacturing process that achieves directional material removal through ion–radical interactions in the plasma, with process outcomes being highly sensitive to plasma chamber operating conditions.
Process behavior is governed by chamber control parameters, including gas flow, RF power, chamber pressure, and temperature, which are collected as in-situ time-series data for monitoring and control.
Since plasma states cannot be fully characterized by process parameters alone, OES is commonly employed to capture active species dynamics and plasma state variations in real time.
Accordingly, multi-channel in-situ time series composed of process parameters and OES serve as a primary information source, and predicting wafer-level etch distributions and spatial non-uniformities from these signals is a key challenge for process monitoring and anomaly detection.

% \section{Dataset and Preprocessing}
\section{Dataset Preparation}

%\subsection{Dataset}
\subsection{Dataset Overview}

% 본 연구에서는 BOSCH Deep Reactive Ion Etching (DRIE) process dataset~\cite{sayyed2025bosch}을 사용하였다.
% 해당 데이터셋은 MEMS 공정을 대상으로 수집되었으며, SF$_6$/C$_4$F$_8$ 기반의 플라즈마 식각 공정을 사용하므로 공정 물리 관점에서 반도체 제조에서 사용되는 plasma-based RIE 공정과 동일한 메커니즘을 따른다.
% 데이터셋은 10장의 wafer로 구성된 lot을 반복 수행한 100회 미만의 식각 공정(run)으로 이루어져 있으며, cleaning 없이 각 lot 내에서 wafer를 연속 처리함으로써 챔버 상태 변화와 공정 drift가 wafer 순서에 따라 누적되도록 설계되었다.
% 각 run에 대해 process parameter 시계열, OES 시계열, 그리고 wafer 내 89개 위치에서 측정된 etch depth로 구성된 wafer-level spatial profile이 제공된다.
% PP, OES, 그리고 wafer map의 대표적인 예시는 Fig.~\ref{fig:data_ex}에 제시되어 있다.
% 모든 공정은 동일한 etch target 조건에서 수행되었으며, 이로 인해 wafer 간 전역적 drift와 wafer 내부의 공간적 비균일성이 동시에 관측된다.

\begin{figure}[t]
    \centering
    % ---------- Row 1 : (a), (b) ----------
    \begin{minipage}[t]{0.3\linewidth}
        \centering
        \includegraphics[width=\linewidth]{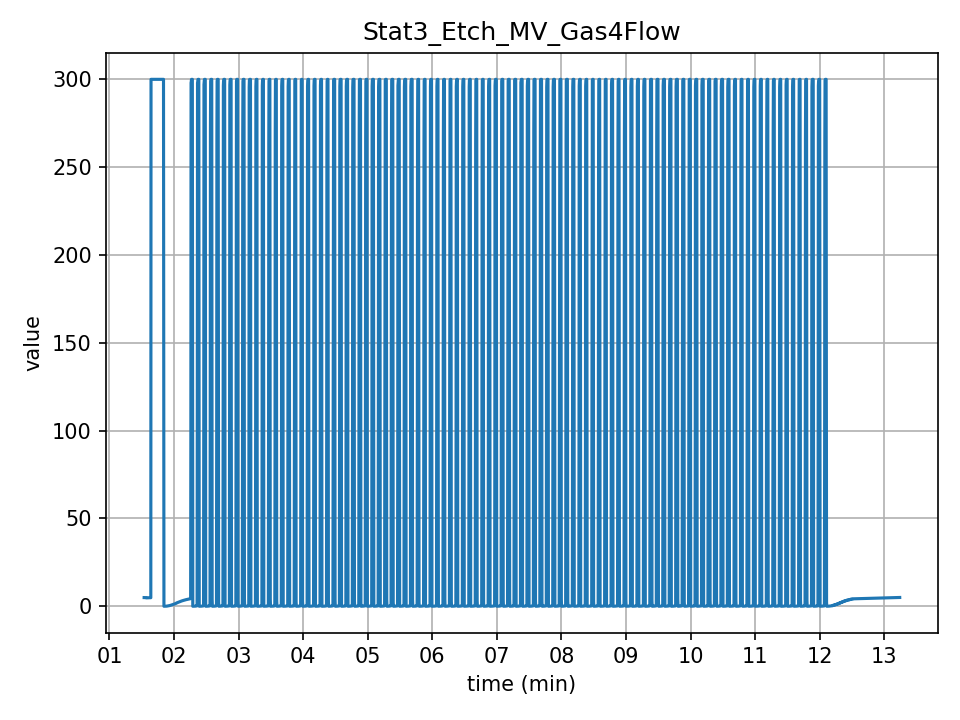}
        \subcaption{}
    \end{minipage}
    \hspace{0.02\linewidth}
    \begin{minipage}[t]{0.3\linewidth}
        \centering
        \includegraphics[width=\linewidth]{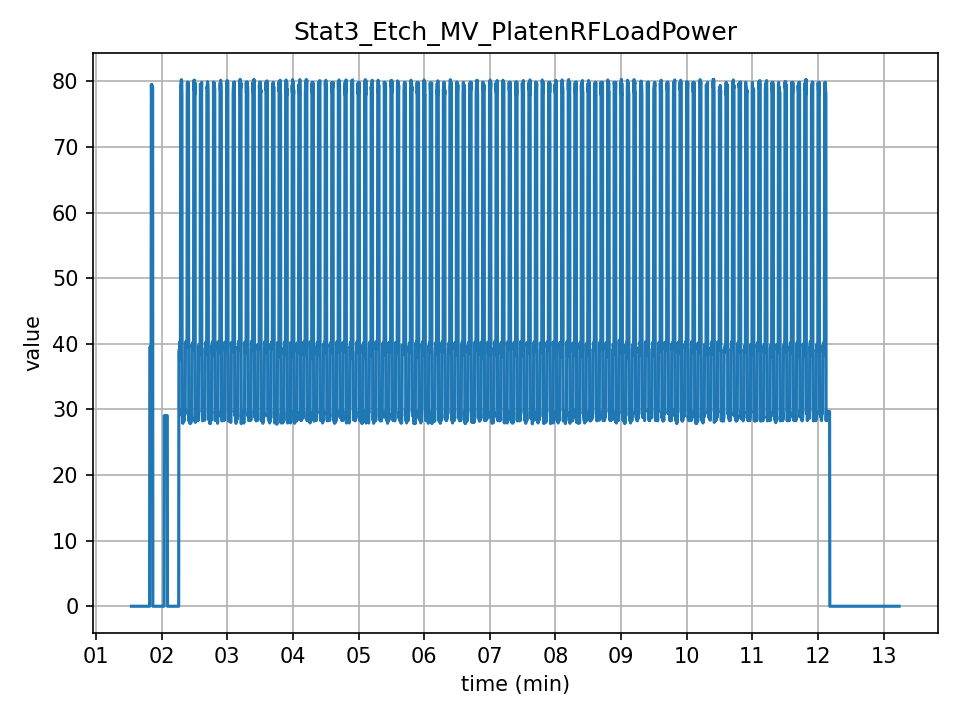}
        \subcaption{}
    \end{minipage}
    \hspace{0.02\linewidth}
    \begin{minipage}[t]{0.3\linewidth}
        \centering
        \includegraphics[width=\linewidth]{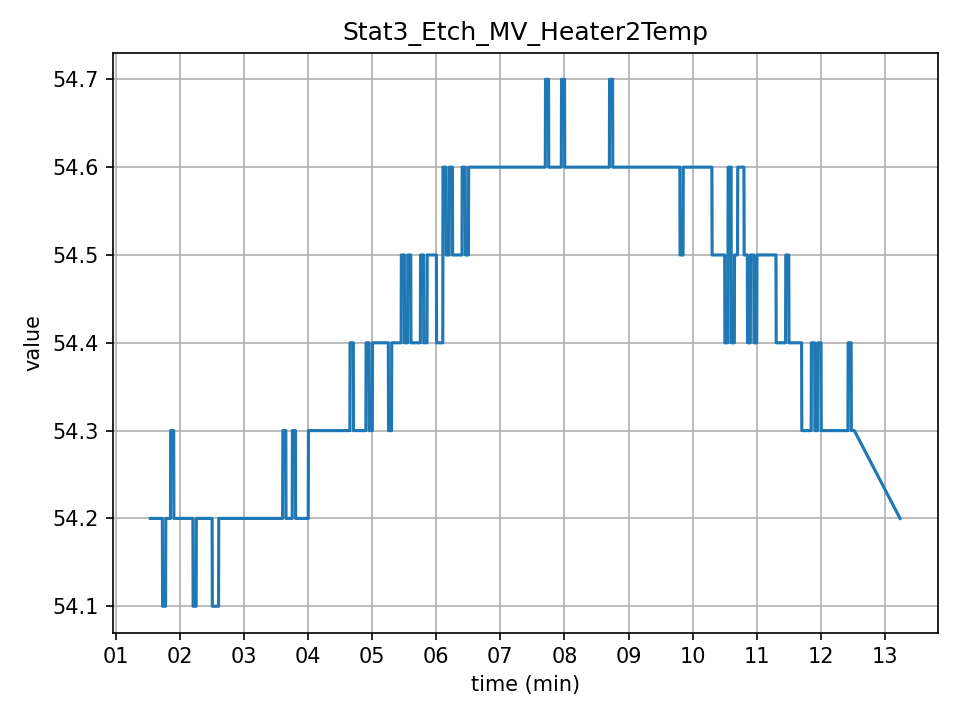}
        \subcaption{}
    \end{minipage}
    % ---------- Row 2 : (c), (d) ----------
    \begin{minipage}[t]{0.6\linewidth}
        \centering
        \includegraphics[width=\linewidth]{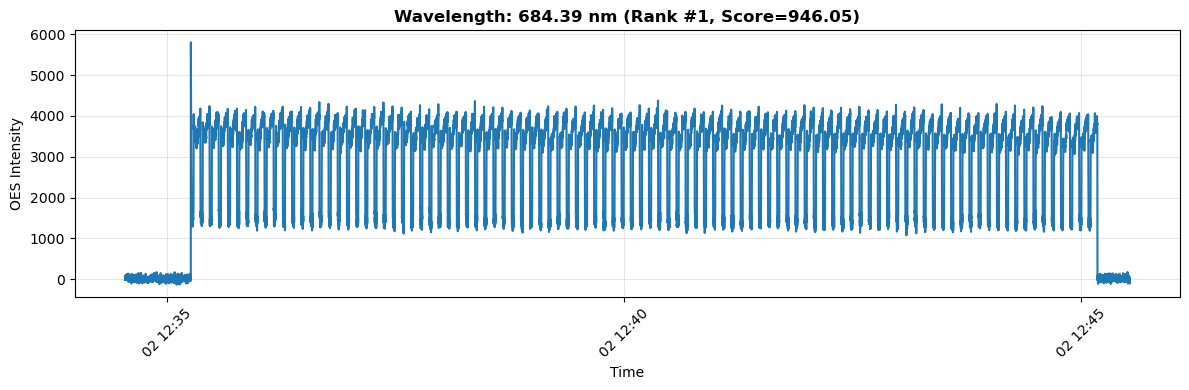}
        \subcaption{}
    \end{minipage}
    \hspace{0.02\linewidth}
    \begin{minipage}[t]{0.3\linewidth}
        \centering
        \includegraphics[width=\linewidth]{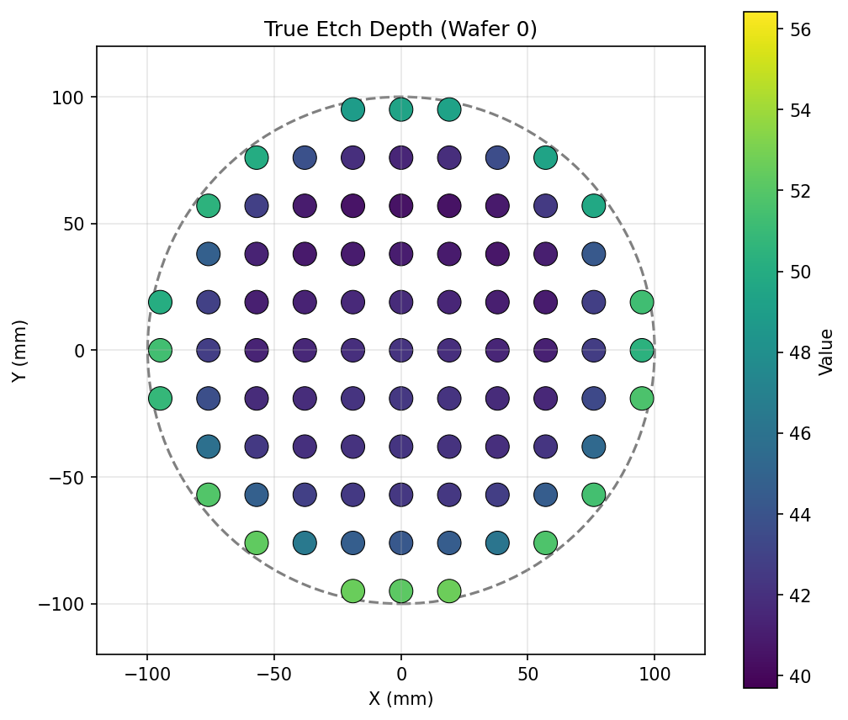}
        \subcaption{}
    \end{minipage}
    \caption{Example in-situ and ex-situ data from the BOSCH DRIE dataset:
    (a) gas flow, (b) RF power, (c) heater temperature, (d) OES intensity, and (e) wafer-level etch depth profile.
    }
    \label{fig:data_ex}
\end{figure}

This study uses the BOSCH Deep Reactive Ion Etching (DRIE) dataset~\cite{sayyed2025bosch}, collected from MEMS fabrication processes employing an SF$_6$/C$_4$F$_8$–based plasma etching scheme.
Although acquired in a MEMS context, the dataset follows the same plasma-based etching mechanisms commonly used in semiconductor manufacturing.
% The dataset comprises fewer than 100 etching runs, each consisting of a lot of 10 wafers processed sequentially without chamber cleaning, thereby inducing cumulative chamber condition changes and process drift along the wafer order within each lot.
The dataset comprises fewer than 100 wafer-level etching runs, organized into multiple lots, each lot consisting of 10 wafers processed sequentially without chamber cleaning.
This sequential processing within each lot induces cumulative chamber condition changes and process drift along the wafer order.
For each run, in-situ process parameter and OES time series are provided together with wafer-level etch depth profiles measured at 89 locations.
Representative examples are shown in Fig.~\ref{fig:data_ex}.
All runs were conducted under identical etch target conditions, allowing simultaneous observation of global wafer-to-wafer drift and intra-wafer spatial non-uniformity.

\subsection{Channel Selection}

% 본 연구에서는 제한된 wafer 수 환경에서의 학습 안정성과 일반화 성능을 고려하여, process parameter(PP) 신호와 OES 스펙트럼으로부터 예측에 유의미한 feature 및 band만을 선별하였다.
% PP feature selection은 전체 공정 주기와 모든 wafer에 대해 값의 분산이 거의 없거나 표준편차가 매우 작은 채널을 제거하는 방식으로 수행하였다.
% OES band selection은 파장별 intensity의 시간적 변동 특성을 기준으로 수행하였다.
% 각 파장에 대해 전체 공정 주기 동안의 intensity 표준편차와 시간 차분의 평균 절대값을 함께 고려하여 점수를 산출하였으며, wafer 간 변동에 강건하도록 해당 통계량을 wafer 기준으로 median aggregation하였다.
% 이후 점수가 높은 상위 K개의 파장을 선택하고, 인접 파장 간 중복 선택을 방지하기 위해 파장 축 상에서 non-maximum suppression(NMS)을 적용하였다.

To ensure stable learning and generalization under a limited number of wafers, informative channels are selected from process parameter signals and OES data.
Process parameter channels whose values exhibit negligible variance or extremely small standard deviation across the entire process duration and all wafers are removed, as they contribute little to prediction performance.
OES channel selection is based on temporal intensity variability.
Each wavelength is scored using the standard deviation and mean absolute temporal difference over the process cycle, with wafer-wise median aggregation applied for robustness.
The top $K$ wavelengths are selected, and non-maximum suppression (NMS) is applied along the wavelength axis to avoid redundant selection of adjacent bands.

\subsection{Dataset Conditioning}

% Feature 및 band selection 이후, in-situ 시계열에서 실제 식각 공정 구간만을 추출하고 PP와 OES 데이터의 시간 축을 정합하여 모델 입력을 구성하였다.
% 공정 구간은 gas flow와 RF power 등 플라즈마 챔버 구동을 직접 반영하는 process parameter 신호를 기준으로 정의하였으며, 해당 구간에 맞추어 OES 시계열을 시간 축 상에서 정렬하였다.
% 정렬된 PP 및 OES 시계열은 모델 입력 길이에 맞추기 위해 시간 축에서 재표본화한 뒤, time step 단위로 일대일 대응되도록 구성하였다.
% 이 과정을 통해 각 wafer에 대해 고정 길이의 동기화된 다채널 in-situ 시계열 입력을 생성하였고, 이에 대응되는 89-point etch depth를 label로 사용하였다. Missing data를 제외한 총 88개의 wafer를 실험에 사용하였다.

After feature and band selection, the active etching phase, during which plasma-assisted etching is performed, is identified from the in-situ time series using process parameter signals that directly reflect chamber operation, such as gas flow and RF power.
The OES time series is then temporally aligned to this phase.
The aligned process parameter and OES signals are resampled to match the fixed model input length and organized to ensure one-to-one correspondence at each time step. This procedure yields fixed-length, synchronized multi-channel in-situ inputs for each wafer, with the corresponding 89-point etch depth spatial profiles used as labels.
After excluding wafers with missing data, a total of 88 wafers are used for evaluation.

\section{Proposed Model Architecture}

% 본 연구에서는 식각 공정 중 수집된 in-situ 시계열 데이터인 process parameter(PP) 신호와 optical emission spectroscopy(OES) 신호로부터 wafer의 89-point etch depth 공간 분포를 예측하는 Time-LLM 기반 공간 회귀 모델을 제안한다.
% 전체 모델 구조는 Fig.~\ref{fig:model}에 제시되어 있다.
% The overall processing pipeline follows the standard Time-LLM framework described in Section~\ref{sec:timellm}, with task-specific modifications introduced for spatial regression.
% 제안 모델은 기존 Time-LLM 구조를 기반으로 하되, 시계열 forecasting을 위해 설계된 입력·출력 구조를 wafer-level 공간 프로파일 회귀 문제에 맞게 재구성하였다.
% 구체적으로, input embedding, prompt-as-prefix(PaP), output projection, 그리고 손실 함수를 수정하여 in-situ 시계열로부터 wafer의 공간 분포를 직접 예측하도록 확장하였다.
% 반면, Time-LLM의 핵심 설계 원칙인 channel-independent 시계열 표현 학습 구조는 그대로 유지하였다.

\begin{figure}[t]
  \centering
  \includegraphics[width=\linewidth]{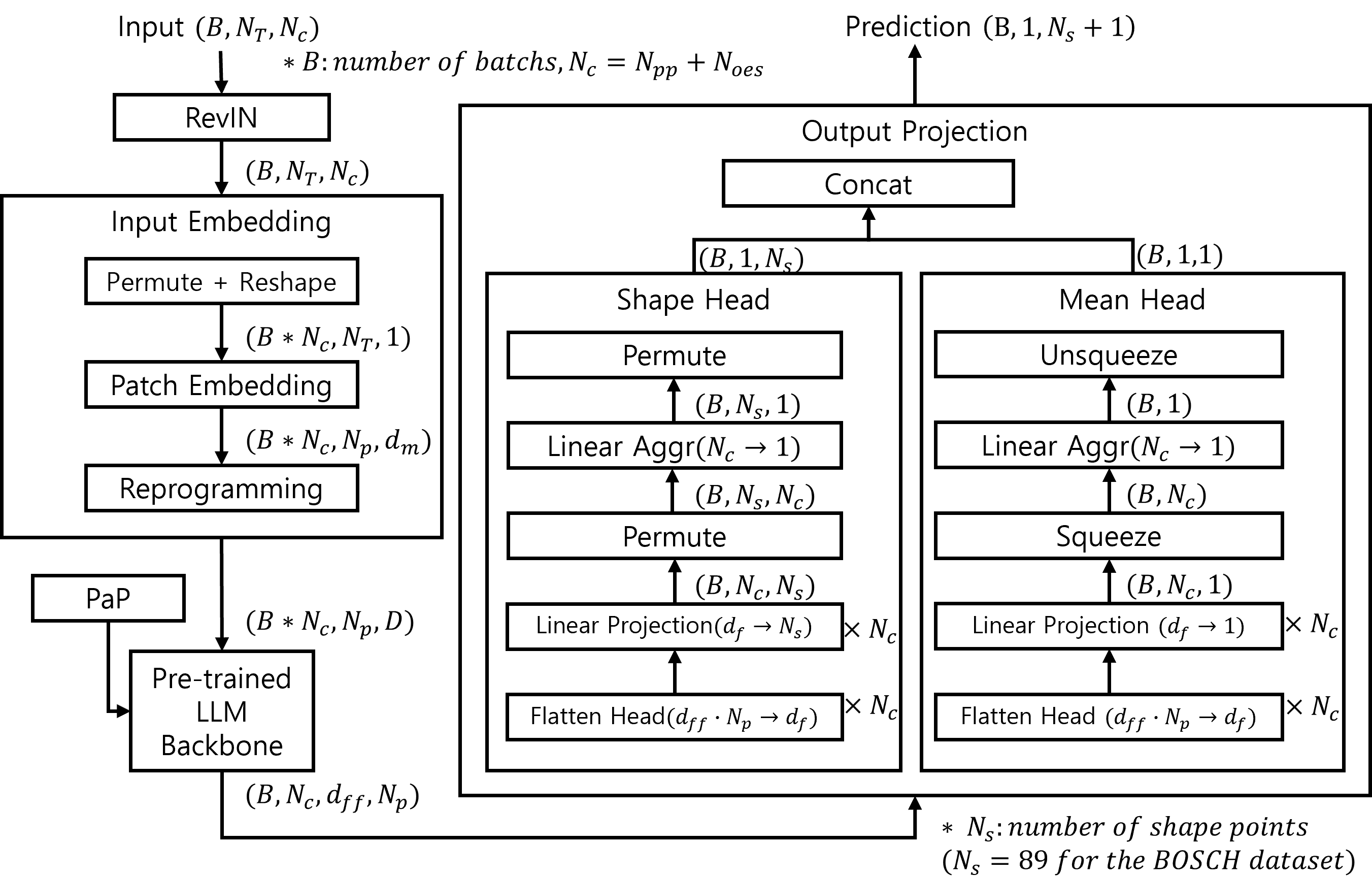}
  %\caption{Proposed model architecture ($N_{s}$ = 89 shape points for the BOSCH DRIE dataset).
  \caption{Proposed model architecture.}
  \label{fig:model}
\end{figure}

We propose a Time-LLM–based spatial regression model that predicts wafer-level etch depth distributions at 89 spatial locations from in-situ time-series data, including process parameters and OES signals.
The overall architecture is illustrated in Fig.~\ref{fig:model}.
The proposed framework follows the standard Time-LLM pipeline described in Section~\ref{sec:timellm}, while introducing task-specific modifications to enable wafer-level spatial regression.
% Our approach extends the standard Time-LLM pipeline with task-specific modifications for wafer-level spatial regression, as described in Section~\ref{sec:timellm}.
In particular, the input embedding, prompt-as-prefix (PaP), output projection, and loss function are redesigned to map temporal in-situ signals directly to wafer-level spatial profiles, rather than to future time steps.
The core design principle of Time-LLM---channel-independent time-series learning with a frozen LLM backbone---is preserved.
% The core design principle of Time-LLM, namely channel-independent time-series representation learning with a frozen LLM backbone, is preserved

\subsection{Input Embedding}

% Input embedding 단계는 식각 공정의 in-situ 시계열 데이터를 patch 단위로 분할하고, 각 patch를 text prototype으로 변환하여 시계열 데이터를 LLM의 embedding 공간에 정렬하는 역할을 수행한다.
% 입력 시계열은 feature/band selection을 통해 선별된 $N_{pp}$개의 process parameter 신호와 $N_{oes}$개의 OES band로 구성되며, 전체 공정 주기에 대해 resampling된 $N_T$개의 시계열 샘플을 포함한다.
% 따라서 wafer 하나에 대한 입력은 $(N_c, N_T)$ 형태의 텐서로 표현되며, $N_c = N_{pp} + N_{oes}$이다.

The input embedding stage transforms in-situ process time series into representations aligned with the LLM embedding space by partitioning the signals into patches and reprogramming each patch as a text prototype.
The input consists of $N_{pp}$ process parameter channels and $N_{oes}$ OES bands selected through channel selection, resampled to $N_T$ time steps over the entire etching process.
Accordingly, each wafer is represented as a tensor of shape $(N_c, N_T)$, where $N_c {=} N_{pp} {+} N_{oes}$.

% 기존 Time-LLM은 시계열 forecasting 문제를 대상으로, 동일한 수치 도메인에서 미래 시계열을 예측하도록 입력과 출력 구조가 설계되었다.
% 반면 본 연구에서는 wafer 하나의 전체 공정 시계열을 입력으로 사용하여, wafer-level 2-D 공간 프로파일을 직접 회귀 예측하는 문제를 다룬다.
% 이에 따라 본 연구에서는 Time-LLM의 reprogramming 개념을 유지하되, 식각 공정 시계열과 공간 회귀 태스크에 맞도록 input embedding 구조를 재구성하였다. 
% Input embedding은 normalization을 거친 input에 대해 patch embedding, 그리고 patch reprogramming의 순서로 수행된다.

While the Time-LLM is designed for forecasting future values within the same numerical domain,
this work uses the entire process time series of a wafer as input and directly regresses a wafer-level spatial profile.
To this end, the input embedding structure is reconfigured while preserving the core reprogramming concept of Time-LLM.
Specifically, the embedding pipeline applies normalization followed by patch embedding and patch reprogramming, enabling the temporal process signals to be mapped to spatial regression outputs.

\subsubsection{Normalization}

% 입력 시계열의 normalization은 채널 간 스케일 차이를 완화하고 학습 안정성을 확보하기 위해 수행된다.
% 본 연구에서는 Time-LLM에서 사용된 RevIN 기반 normalization을 동일하게 적용하지만, forecasting 문제와 달리 출력 도메인이 입력과 다른 회귀 문제의 특성을 고려하여 출력 단계에서는 denormalization을 수행하지 않는다.
% 기존 Time-LLM은 입력과 출력의 scale이 동일하므로 출력에서 원래 스케일을 복원하지만, 본 연구에서는 입력이 공정 신호(PP 및 OES)이고 출력이 wafer etch depth이므로 이러한 과정이 필요하지 않다.
% 입력 시계열은 wafer별·채널별로 전체 공정 주기에 대한 평균과 표준편차를 기준으로 정규화되며, 해당 scale 정보는 Prompt-as-Prefix(PaP)에 포함된 통계 정보로 LLM에 전달된다.

Normalization is applied to the input time series to mitigate scale disparities across channels and to improve training stability.
We adopt the RevIN-based normalization scheme used in Time-LLM; however, unlike forecasting tasks, denormalization is not applied at the output stage, reflecting the cross-domain nature of the regression task.
Specifically, the inputs consist of process parameter and OES signals, whereas the outputs are wafer-level etch depths, for which scale recovery is unnecessary.
The input signals are normalized on a per-wafer, per-channel basis using statistics computed over the entire etching process.
These scale statistics are provided to the LLM through the Prompt-as-Prefix (PaP), allowing the model to retain scale-related information while operating on normalized inputs.

% 입력 시계열의 scale 정보는 Prompt-as-Prefix(PaP)에 포함된 통계 정보로 LLM에 전달된다. Normalization은 wafer별, 채널 $i$별로 전체 공정 주기에 대한 평균 $\mu^{(i)}$와 표준편차 $\sigma^{(i)}$를 이용해 다음과 같이 수행된다:
% \begin{equation*}
%     \tilde{X}^{(i)}(t) = \frac{X^{(i)}(t)-\mu^{(i)}}{\sigma^{(i)}}
%     \label{eq:norm}
% \end{equation*}

\subsubsection{Patch Embedding}  

% Patch embedding은 정규화된 입력 시계열을 시간 축을 따라 국소적인 시계열 단위의 patch로 분할하고, 각 patch를 임베딩 벡터로 변환하는 과정이다.
% 기존 Time-LLM은 긴 시계열을 여러 개의 forecasting sample로 분할하는 sliding-window 방식으로 학습 데이터를 구성한다.
% 반면 본 연구에서는 wafer 하나의 전체 공정 시계열이 하나의 학습 샘플에 해당하므로, 전체 공정 주기 $N_T$를 단일 입력으로 사용한다.
% 즉, forecasting 관점에서의 sequence length를 $N_T$로 설정하고, 이 시계열을 내부적으로 patch로 분할하여 LLM이 처리 가능한 token sequence를 구성한다.
% 각 PP 및 OES 채널은 길이 $L_p$, stride $S$의 patch로 분할되며, 시작과 종료 구간을 포함하기 위해 boundary patch를 추가한다.
% 이에 따라 채널별 patch 개수는 $N_p{=}\left\lfloor {(N_T - L_p)}/{S} \right\rfloor + 2$로 정의된다.
% 각 patch는 선형 변환을 통해 $L_p$ 길이의 시계열에서 $d_m$차원 임베딩으로 projection되며, 이는 공정 시계열의 국소적인 시간적 패턴을 요약한 표현에 해당한다.

Patch embedding converts the normalized input time series into a sequence of local temporal representations by partitioning each channel into patches and projecting them into fixed-dimensional embeddings.
In contrast to Time-LLM, which generates multiple training samples via sliding windows for forecasting, this work treats the entire process time series of a wafer as a single input of length $N_T$, which is used as the sequence length in the Time-LLM framework.
The input sequence is then internally segmented into patches to form an LLM-compatible token sequence.
Each process parameter and OES channel is divided into patches of length $L_p$ with stride $S$, including boundary patches at the sequence ends, yielding $N_p {=} \left\lfloor {(N_T - L_p)}/{S} \right\rfloor {+} 2$ patches per channel.
Each patch is linearly projected into a $d_m$-dimensional embedding that summarizes local temporal patterns in the process signals.
\begin{equation*}
    \mathbf{X}^{(i)} \in \mathbb{R}^{N_p \times L_p} \xrightarrow[]{\text{Linear}} \mathbf{\hat{X}}^{(i)} \in \mathbb{R}^{N_p \times d_m}
    \label{eq:patch_embedding}
\end{equation*}
  
\subsubsection{Patch Reprogramming}

% Patch reprogramming은 patch embedding으로 생성된 시계열 token을 LLM 입력 공간에 적합한 word token 표현으로 변환하는 단계이다.
% 본 연구에서는 Time-LLM에서 제안된 input reprogramming 방식을 그대로 적용하였다.
% 이는 해당 방식이 시계열 패턴을 LLM의 사전학습된 언어 표현 공간과 효과적으로 정렬할 수 있음이 이미 검증되었으며, 본 연구의 목표가 reprogramming 메커니즘 자체의 변경이 아니라 시계열-to-공간 회귀 문제로의 확장에 있기 때문이다.
% 따라서 multi-head cross attention과 선형 변환을 통해 $d_m$차원의 시계열 token을 LLM backbone의 hidden dimension $D$ 차원으로 사상하는 구조를 유지하였다.
% Attention head의 개수를 $K$라 할 때, 각 head의 차원은 $d=\lfloor d_m/K \rfloor$이다.

Patch reprogramming maps the time-series tokens produced by patch embedding into word-token representations compatible with the LLM input space.
In this work, we directly adopt the input reprogramming mechanism proposed in Time-LLM, as it has been shown to effectively align temporal patterns with the pretrained linguistic embedding space.
Since the objective of this study is to extend Time-LLM from temporal forecasting to time-series–to–spatial regression, rather than to modify the reprogramming mechanism itself, the original structure is preserved.
Specifically, the $d_m$-dimensional patch tokens are projected to the LLM backbone hidden dimension $D$ using multi-head cross-attention followed by a linear transformation.
With $K$ attention heads, each head has dimensionality $d=\lfloor d_m/K \rfloor$:
\begin{equation*}
    \mathbf{X}^{(i)} \in \mathbb{R}^{N_p \times d_m}
    \xrightarrow[]{\text{Attention}} 
    \{\mathbf{\hat{Z}}^{(i)}_k \in \mathbb{R}^{N_p \times d}\}_{k=1}^K
    \label{eq:attention}
\end{equation*}
\begin{equation*}
    \mathrm{Concat}(\mathbf{\hat{Z}}^{(i)}_1, \dots, \mathbf{\hat{Z}}^{(i)}_K)
    \xrightarrow[]{\text{Linear}}
    \mathbf{\hat{O}}^{(i)} \in \mathbb{R}^{N_p \times D}
    \label{eq:patch_reprog}
\end{equation*}
% Reprogramming된 $N_p$개의 token sequence는 Prompt-as-Prefix(PaP)와 함께 LLM에 입력된다.
% Time-LLM의 channel-independent 설계를 계승하여, 각 입력 채널은 독립적으로 LLM 추론을 수행한다.
The resulting sequence of $N_p$ reprogrammed tokens is concatenated with the Prompt-as-Prefix (PaP) and fed into the LLM.
Consistent with the channel-independent design of Time-LLM, each input channel is processed independently by the LLM during inference.

% 구현 측면에서는 채널 차원을 batch 차원에 포함시켜 병렬 연산을 수행하며, 이에 따라 실제 LLM 입력의 batch size는 $\text{batch size} \times N_c$가 된다.
% 따라서 실제 batch size는 $\text{batch size} \times N_c$가 된다.

\subsection{Output Projection}

% Output projection 단계는 LLM의 출력 표현을 wafer의 89개 공간 측정 지점에 대한 etch depth 분포로 사상한다.
% 본 연구에서는 wafer-level etch depth를 전역 평균(mean)과 공간 분포 형태(shape)로 분리하여 예측한다.
% Shape는 wafer 내 평균값을 제거한 공간적 잔차로 정의되며, center–edge 차이, 링 패턴, 비대칭성과 같은 공간 분포의 형태적 특성만을 포함한다.
% 반면 mean은 wafer 전체의 절대적인 식각 수준을 나타내며, 공정 조건 변화나 챔버 상태 열화에 따른 전역적 drift를 주로 반영한다.
% 두 성분은 물리적 의미와 변동 메커니즘이 상이하므로, 본 연구에서는 이를 분리하여 예측하도록 출력 구조를 설계하였다.

The output projection stage maps the LLM output representations to wafer-level etch depth distributions at 89 spatial locations.
To capture distinct sources of variation, the wafer-level etch depth is decomposed into a global mean component and a spatial shape component.
The shape is defined as the spatial residual after removing the wafer-wise mean, representing morphological characteristics such as center–edge variation, ring patterns, and asymmetry.
In contrast, the mean reflects the absolute etch level across the wafer and captures global drift induced by process condition changes or chamber aging.
Given their different physical interpretations and variation mechanisms, these two components are predicted separately.

% 기존 Time-LLM은 시계열 forecasting을 목적으로 하여, prediction horizon과 output channel 차원으로 출력을 정의한다.
% 반면 본 연구에서는 시계열-to-공간 회귀 문제를 다루기 위해 prediction horizon을 1로 설정하고, output channel을 wafer 공간 좌표에 대응되는 89개의 shape 값과 1개의 mean 값으로 재구성하였다.
% 형식적으로는 단일 시점의 다채널 출력을 예측하는 구조이나, 실제로는 wafer-level 공간 좌표에 대한 etch depth를 회귀하는 문제에 해당한다.
% % Time-LLM의 출력 채널은 서로 독립적으로 처리되므로, 공간적 상관관계는 직접적으로 고려되지 않는다.

Compared with Time-LLM, which defines outputs in terms of prediction horizons and output channels for temporal forecasting,
this work reformulates the output for time-series–to–spatial regression by setting the prediction horizon to one and redefining the output channels as 89 spatial shape values and a single mean value.
Although the formulation corresponds to a single-step multi-channel output, it effectively performs regression over wafer-level spatial coordinates.

% Output projection은 flatten layer, linear projection layer, 그리고 linear aggregation layer의 세 단계로 구성된다.
% Flatten 및 per-channel projection 구조는 Time-LLM의 channel-independent 설계를 그대로 계승하되, forecasting 출력을 공간 회귀 출력으로 변환하도록 projection 및 aggregation 구조를 새롭게 설계하였다.
% % Output projection은 shape 예측과 mean 예측을 위한 두 개의 독립적인 head로 구성된다.

The output projection consists of three stages: a flatten layer, a linear projection layer, and a linear aggregation layer.
While the flattening and per-channel projection follow the channel-independent design of Time-LLM, the projection and aggregation structures are redesigned to transform temporal representations into spatial regression outputs.

\subsubsection{Flatten Layer}

% Flatten layer는 LLM의 출력 표현을 채널별 1차원 벡터로 변환하여 이후 projection 단계의 입력으로 사용한다.
% 이 단계는 기존 Time-LLM의 설계를 그대로 따른다.
% LLM은 각 patch에 대해 hidden dimension $D$의 출력을 생성하며, 이 중 prefix에 해당하는 구간을 제외하고 $d_{ff}$차원만을 patch별 feature로 사용한다.
% 결과적으로 채널 $i$에 대한 출력은 $\mathbb{R}^{N_p \times d_{ff}}$ 형태가 되며, flatten layer에서는 이를 하나의 벡터 $\mathbb{R}^{d_f}$ ($d_f = N_p \cdot d_{ff}$)로 변환한다. 
% 이 연산은 shape head와 mean head에서 동일하게 적용된다.

% The flatten layer converts the LLM output representations into channel-wise one-dimensional vectors for subsequent projection. This stage follows the original Time-LLM design. Specifically, the LLM produces hidden representations of dimension $D$ for each patch, from which only the $d_{ff}$-dimensional features corresponding to non-prefix tokens are retained. As a result, the output for channel $i$ has shape $\mathbb{R}^{N_p \times d_{ff}}$, which is flattened into a single vector in $\mathbb{R}^{d_f}$, where $d_f = N_p \cdot d_{ff}$. This operation is applied identically to both the shape and mean heads.

The flatten layer converts the LLM output representations into channel-wise one-dimensional vectors for subsequent projection, consistent with the baseline Time-LLM architecture.
The LLM produces hidden representations of dimension $D$ for each patch, from which the $d_{ff}$-dimensional features excluding the prefix tokens are retained.
Consequently, the output for each channel has shape $\mathbb{R}^{N_p \times d_{ff}}$ and is flattened into a vector in $\mathbb{R}^{d_f}$, where $d_f{=}N_p \cdot d_{ff}$. 
This operation is applied to both the shape and mean heads.
  
\subsubsection{Linear Projection Layer}

% Linear projection layer는 채널별로 flatten된 $d_f$차원의 출력 벡터를 wafer 공간 좌표에 대응되는 예측 값으로 사상한다.
% 기존 Time-LLM에서는 해당 표현을 미래 시점의 prediction horizon으로 mapping하였으나, 본 연구에서는 이를 89개의 spatial point(shape)와 1개의 scalar mean 값으로 mapping하도록 수정하였다.
% 채널 $i$에 대해 shape head와 mean head는 각각 독립적인 선형 변환을 수행하여, 채널별 shape 벡터 $\hat{\mathbf{S}}^{(i)} \in \mathbb{R}^{89}$와 mean 값 $\hat{m}^{(i)} \in \mathbb{R}$를 생성한다.
The linear projection layer maps the channel-wise flattened vectors of dimension $d_f$ to predictions defined over wafer spatial coordinates and a global mean.
While the Time-LLM projects these representations to future prediction horizons for temporal forecasting, this work reformulates the projection to produce 89 spatial shape values and a single scalar mean value.
For each input channel $i$, the shape head and mean head apply independent linear transformations to generate a channel-specific shape vector $\hat{\mathbf{S}}^{(i)}$ and a mean estimate $\hat{m}^{(i)}$:
\begin{equation*}
    \hat{\mathbf{S}}^{(i)}_f \in \mathbb{R}^{d_f} \xrightarrow[]{\text{Linear}_s} \hat{\mathbf{S}}^{(i)} \in \mathbb{R}^{89}
    \label{eq:lin_shape}
\end{equation*}
\begin{equation*}
    \hat{\mathbf{M}}^{(i)}_f \in \mathbb{R}^{d_f} \xrightarrow[]{\text{Linear}_m} \hat{m}^{(i)} \in \mathbb{R}
    \label{eq:lin_mean}
\end{equation*}
% 이 단계에서는 각 입력 채널이 제공하는 공정 정보가 공간 분포와 전역 평균 예측에 어떻게 기여하는지를 개별적으로 학습한다.
This design enables the model to learn how the process information encoded in each input channel contributes independently to spatial distribution and global mean estimation.

\subsubsection{Linear Aggregation Layer}

% Linear aggregation layer에서는 채널별로 생성된 shape 및 mean 예측값을 종합하여 최종 wafer-level 예측 결과를 도출한다.
% 기존 Time-LLM은 입력과 출력이 동일한 도메인에 속하므로 채널별 예측값을 그대로 사용하지만, 본 연구에서는 입력이 in-situ 공정 시계열이고 출력이 wafer 공간 프로파일로 서로 다른 도메인에 속한다.
% 따라서 여러 입력 채널의 정보를 통합하는 aggregation 과정이 필요하다.

The linear aggregation layer integrates the channel-wise shape and mean predictions to produce the final wafer-level outputs.
Unlike the Time-LLM, where inputs and outputs reside in the same domain and channel-wise predictions can be used directly, the present task maps in-situ process time series to wafer-level spatial profiles.
This cross-domain setting necessitates explicit aggregation of information across multiple input channels.

% Shape head에서는 각 spatial point $j$에 대해 채널별 예측값을 선형 결합하여 aggregation을 수행하며, 모든 spatial point에 대해 동일한 weight를 공유한다.
% 이는 위치에 따라 다른 해석 규칙을 학습하는 대신, 공간 전반에 대해 일관된 해석을 적용한다는 가정을 반영한 설계이다.
% Spatial point별로 독립적인 aggregation을 둘 경우 파라미터 수가 급격히 증가하여, 제한된 데이터 환경에서 overfitting 위험이 커질 수 있다.

For the shape head, channel-specific predictions are linearly combined at each spatial point using shared aggregation weights across all locations.
This design enforces a consistent interpretation of channel contributions over the wafer, 
while avoiding a location-dependent aggregation that would substantially increase the number of parameters and exacerbate overfitting under limited data:
\begin{equation*}
    \hat{s}_j^{raw} = \sum\nolimits_{i=1}^{N_c} w_i^{\text{shape}} \hat{s}_j^{(i)}, \quad j=1,...,89
    \label{eq:shape_aggr}
\end{equation*}
% Shape 예측 결과에는 mean bias가 포함될 수 있으므로, 최종 출력에서는 전체 spatial point에 대한 평균을 제거하여 순수한 공간 분포 형태만을 남긴다.
Since the aggregated shape prediction may contain a residual mean bias, the final shape output is obtained by removing the spatial average, yielding a zero-mean spatial profile:
\begin{equation*}
    \hat{s}_j = \hat{s}_j^{raw} - \mu_s, \quad \mu_s = \frac{1}{89} \sum\nolimits_{j=1}^{89} \hat{s}_j^{raw}
    \label{eq:bias_remove}
\end{equation*}
% Mean head에서는 채널별 mean 예측값을 선형 결합하여 하나의 scalar mean 값을 도출하며, shape head와는 독립적인 aggregation 계층으로 구현된다.
The mean head aggregates channel-wise mean estimates through an independent linear combination to produce a single scalar mean value:
\begin{equation*}
    \hat{m} = \sum\nolimits_{i=1}^{N_c} w_i^{\text{mean}} \hat{m}^{(i)}
    \label{eq:mean_aggr}
\end{equation*}

% 앞서 언급한 바와 같이 output projection 단계에서는 denormalization을 적용하지 않으며, 출력값과 ground-truth label 역시 정규화하지 않는다.
% 출력값을 정규화할 경우 wafer 간 공간 비균일성의 절대적인 크기 정보가 제거되어, 상대적으로 심각한 공간 비균일성을 보이는 wafer를 구분하기 어렵게 된다.
% 또한 공정 조건 변화나 챔버 상태 열화에 따른 전역적 drift 정보가 소실될 수 있으므로, 본 연구에서는 etch depth 값을 원래 스케일 그대로 사용하였다.
As discussed earlier, no denormalization is applied at the output stage, and both predictions and ground-truth labels remain in their original scale.
Normalizing the outputs would remove information about the absolute magnitude of spatial non-uniformity across wafers 
and obscure global drift induced by process condition changes or chamber degradation.
Therefore, etch depth values are preserved in their physical scale throughout training and evaluation.

\subsection{Prompt-as-Prefix (PaP)}

% Prompt-as-Prefix(PaP)는 시계열 patch 입력 앞에 위치하는 자연어 프롬프트로, LLM이 이후 입력되는 시계열 토큰을 어떤 과제로 해석하고 변환해야 하는지를 명시적으로 지시한다.
% 본 연구에서는 Time-LLM에서 제안된 PaP 구조를 그대로 유지하되, 식각 공정의 wafer-level 공간 프로파일 예측 과제에 맞게 프롬프트 내용을 수정하였다.
% PaP는 dataset description, task instruction, 그리고 input statistics로 구성된다. Dataset description은 식각 공정 데이터의 구성과 센서 특성을 반영하도록 재정의하였으며, task instruction은 기존 forecasting 과제 대신 입력된 공정 시계열로부터 wafer의 89-point spatial etch profile을 예측하도록 수정하였다.
% Input statistics는 입력 시계열의 trend 및 lag 정보를 요약한 것으로, 공정 시계열의 시간적 특성을 LLM에 전달하기 위해 기존 Time-LLM 설정을 그대로 적용하였다. 
% 본 논문에서 사용한 PaP의 예시는 Fig.~\ref{fig:prompt}에 제시되어 있다.
Prompt-as-Prefix (PaP) is a natural language prefix that conditions the LLM on how to interpret the subsequent time-series patch tokens.
We adopt the PaP mechanism from Time-LLM and tailor its content to wafer-level spatial etch profile prediction in plasma etching processes.
The PaP comprises a dataset description, a task instruction, and input statistics.
The dataset description is customized to reflect etching process data and sensor modalities, while the task instruction is reformulated to predict an 89-point wafer-level spatial etch profile from in-situ process time series.
The input statistics, which summarize temporal characteristics such as trends and lag information, follow the standard Time-LLM formulation.
An example PaP is provided in Fig.~\ref{fig:prompt}.
\begin{figure}[t]
    \fbox{%
    \begin{minipage}{0.95\linewidth}
    \small
    \textbf{Dataset description:}
        Semiconductor silicon etch process monitoring data.
        The dataset contains $N_c$ sensor measurements sampled over $N_T$ timesteps during a wafer fabrication etch process.
        Sensors include plasma parameters (RF power, gas flow, pressure,
        temperature) and optical emission spectroscopy (OES) intensities at various wavelengths.
        The task is to predict the spatial etch depth uniformity distribution across 89 measurement points on the wafer surface.
        
    \vspace{0.5em}
    \textbf{Task instruction:}
        Predict the spatial etch profile at 89 wafer positions given $N_c$ process sensor channels of $N_T$ timesteps
    \end{minipage}}
    \caption{Prompt-as-Prefix (PaP) used in the Time-LLM configuration for wafer-level spatial etch profile prediction.}
%    \caption{PaP used for spatial etch profile prediction.}   
    
    \label{fig:prompt}
\end{figure}

\subsection{Loss Function}

% 본 연구에서는 wafer-level etch depth를 89-point spatial shape와 wafer 전체 mean으로 분리하여 예측하므로, 손실 함수 역시 두 성분의 예측 오차를 각각 고려한다.
% Shape 예측 오차는 89개 spatial point에 대한 평균제곱오차(MSE)로, mean 예측 오차는 예측된 mean 값과 실제 mean 값 간의 MSE로 계산한다.
% 최종 손실은 두 항의 가중합으로 정의되며, 가중치 $\lambda$는 shape와 mean 예측 오차의 상대적 기여도를 조절한다.
Since the wafer-level etch depth is predicted as an 89-point spatial shape and a global mean, the loss function accounts for errors in both components.
The shape loss is defined as the mean squared error (MSE) over the 89 spatial points, with the mean loss computed analogously for the wafer-level mean.
The overall objective is formulated as a weighted sum of the two terms, where the weighting factor $\lambda$ controls their relative contributions:
\begin{equation*}
  \mathcal{L} = || \mathbf{\hat{s}} - \mathbf{s} ||^2_2 + \lambda(\hat{m}-m)^2  
  \label{eq:loss}
\end{equation*}
% 손실 계산에는 정규화되지 않은 prediction 값과 ground-truth label을 그대로 사용하여, wafer 간 공간 비균일성의 절대적인 크기와 전역적 drift 정보를 보존한다.
% 한편, $\lambda$를 학습 가능한 파라미터로 둘 경우 특정 분기가 과도하게 지배적인 퇴화해가 발생할 수 있으므로, 본 연구에서는 $\lambda$를 고정된 하이퍼파라미터로 설정하고 validation을 통해 조정하였다.
Loss computation is performed directly on unnormalized predictions and ground-truth labels to preserve the absolute magnitude of spatial non-uniformity across wafers as well as global drift effects.
To avoid degenerate solutions in which one branch dominates optimization, $\lambda$ is treated as a fixed hyperparameter and selected via validation rather than being learned during training.

\section{Experiments}

% 본 절에서는 제안한 모델의 성능을 shape MSE, mean MSE, and etch depth MAE를 기준으로 평가한다.
% 입력 시계열의 시간 해상도 및 patch 구성, OES 입력 채널 수, 그리고 backbone LLM의 모델 용량이 wafer-level spatial profile 예측 성능에 미치는 영향을 분석한다.
% wafer-level 공간 프로파일을 직접 예측한 선행 연구가 거의 없고 기존 접근들이 단일 스칼라 지표 예측에 국한되어 있어 비교 가능한 기준이 제한적이므로, 단일 평균값으로 wafer를 설명하는 Global Mean Baseline을 비교 대상으로 설정하여 공간 정보를 명시적으로 모델링하는 접근의 효과를 검증한다.
% 모든 실험은 lot 단위 10-fold cross validation으로 수행하였으며, 각 fold에 대한 평균과 표준편차를 함께 보고한다. 실험은 NVIDIA A100 GPU(80GB) 2장을 사용하여 수행하였다.

This section evaluates the proposed model in terms of shape MSE, mean MSE, and etch depth mean absolute error (MAE).
We examine how input temporal resolution and patch configuration, the number of OES channels, and backbone LLM capacity affect wafer-level spatial profile prediction.
Because direct wafer-level spatial profile prediction has been scarcely studied and existing approaches are largely restricted to scalar metrics, suitable baselines are limited.
% We therefore adopt a Global Mean Baseline, which represents each wafer by a single average value, to quantify the benefit of explicitly modeling spatial information.
We therefore adopt a Global Mean Baseline that predicts a constant value equal to the training-set global mean for all wafers, to quantify the benefit of explicitly modeling spatial information.
All experiments are conducted using lot-wise 10-fold cross-validation, reporting the mean and standard deviation across folds, and are performed on two NVIDIA A100 GPUs (80GB).

% \subsubsection{Learning parameter size}

% Requires: \usepackage{booktabs}
\begin{table}[t]
    \centering
    \caption{10-fold cross-validation errors (mean $\pm$ std) for different patch length, stride, and sequence length.}
    \label{tab:patch_seq_ablation}
    \setlength{\tabcolsep}{3pt}
    \renewcommand{\arraystretch}{1.1}
    \begin{tabular}{llcccc}
    \toprule
    \multicolumn{6}{c}{\textbf{MSE (shape)}} \\
    \cmidrule(lr){1-6}
    $L_p$ & $S$ & $N_T{=}3000$ & $N_T{=}6000$ & $N_T{=}9000$ & $N_T{=}12000$ \\
    \midrule
    $32$ & $16$ & $0.42\pm0.33$ & $0.73\pm0.41$ & $0.51\pm0.21$ & $\mathbf{0.18\pm0.10}$ \\
    $48$ & $24$ & $0.57\pm0.35$ & $0.37\pm0.18$ & $0.59\pm0.28$ & $0.44\pm0.44$ \\
    $64$ & $32$ & $0.65\pm0.41$ & $0.53\pm0.40$ & $0.42\pm0.12$ & $0.46\pm0.20$ \\
    \midrule
    \multicolumn{6}{c}{\textbf{MSE (mean)}} \\
    \cmidrule(lr){1-6}
    $L_p$ & $S$ & $N_T{=}3000$ & $N_T{=}6000$ & $N_T{=}9000$ & $N_T{=}12000$ \\
    \midrule
    $32$ & $16$ & $6.26\pm5.92$ & $5.13\pm4.66$ & $2.78\pm2.67$ & $4.16\pm3.67$ \\
    $48$ & $24$ & $4.88\pm5.30$ & $\mathbf{2.28\pm2.00}$ & $7.54\pm5.32$ & $3.99\pm2.65$ \\
    $64$ & $32$ & $4.10\pm2.47$ & $3.70\pm3.29$ & $9.33\pm4.41$ & $4.55\pm7.46$ \\
    \midrule
    \multicolumn{6}{c}{\textbf{MAE (etch)}} \\
    \cmidrule(lr){1-6}
    $L_p$ & $S$ & $N_T{=}3000$ & $N_T{=}6000$ & $N_T{=}9000$ & $N_T{=}12000$ \\
    \midrule
    $32$ & $16$ & $1.88\pm0.76$ & $1.88\pm0.89$ & $\mathbf{1.23\pm0.43}$ & $1.57\pm0.82$ \\
    $48$ & $24$ & $1.73\pm0.92$ & $1.25\pm0.51$ & $2.23\pm0.69$ & $1.61\pm0.56$ \\
    $64$ & $32$ & $1.64\pm0.52$ & $1.53\pm0.71$ & $2.43\pm0.55$ & $1.44\pm0.91$ \\
    \bottomrule
    \end{tabular}
    \label{tab:seq_ps_var}
\end{table}

\subsection{Input Temporal Resolution Sensitivity Analysis}

% 본 실험에서는 입력 시계열의 시간적 해상도($N_T$)와 patch 구성($p/s$)이 wafer-level spatial profile 예측 성능에 미치는 영향을 분석하였다.
% 계산 효율을 고려하여 GPT2 small (6-layer)을 backbone으로 사용하였고, 입력 채널 수는 $N_c{=}73\ (N_{pp}{=}25,\ N_{oes}{=}48)$으로 고정하였다.
% 동일한 공정 구간에 대해 sampling rate를 조절하여 $N_T$를 변화시켰으며, backbone의 context length 제약을 만족하는 설정만을 고려하였다.
% Table~\ref{tab:seq_ps_var}에서 확인할 수 있듯이, $N_T$가 증가할수록 spatial shape MSE는 전반적으로 개선되었고, 
% 특히 $p{=}32/s{=}16$, $N_T{=}12000$ 설정에서 최저 shape MSE가 관측되어 finer temporal resolution이 공간 분포 형태 추론에 효과적임을 확인하였다.
% 반면 mean MSE는 설정에 따른 변동성이 크고 $N_T$ 증가에 따른 일관된 개선은 나타나지 않았다.
% 이는 wafer-level mean이 공정 드리프트와 같은 전역적 요인의 영향을 더 크게 받기 때문으로 해석된다.
% 최종 etch MAE는 전반적으로 shape 예측 성능이 우수한 설정에서 함께 개선되는 경향을 보여, 전체 예측 정확도가 공간 분포 복원에 크게 의존함을 시사한다.
% Patch 구성 측면에서는 과도하게 큰 patch는 국소 패턴을 희석시키고, 지나치게 작은 patch는 context length 제약과 학습 안정성 측면에서 불리하게 작용하여, 중간 수준의 $p/s$ 조합이 성능과 안정성 간 균형을 제공하였다.
We evaluate the effect of input temporal resolution ($N_T$) and patch configuration ($L_p, S$) on wafer-level spatial profile prediction.
Experiments use GPT2-small (6 layers) for computational efficiency, with input channels fixed to $N_c{=}73$ ($N_{pp}{=}25$, $N_{oes}{=}48$).
The temporal resolution $N_T$ is varied by adjusting the sampling rate over the same active etching phase, subject to the backbone context-length constraint.
As summarized in Table~\ref{tab:seq_ps_var}, spatial shape MSE consistently decreases with increasing $N_T$, 
reaching its minimum at $L_p{=}32$, $S{=}16$, and $N_T{=}12000$, which indicates that finer temporal resolution improves spatial pattern recovery. 
In contrast, mean MSE shows no monotonic trend with $N_T$ and varies substantially across configurations, 
reflecting the stronger influence of global factors such as process drift. 
The final etch depth MAE generally improves under configurations that yield lower shape error, highlighting the critical role of accurate spatial reconstruction.
Patch configuration also affects performance: overly large patches obscure local temporal patterns, whereas excessively small patches are limited by context length and training stability.
Overall, intermediate $(L_p, S)$ settings offer the best balance between accuracy and robustness.

\subsection{Input Channel Sensitivity Analysis}

% 본 실험에서는 OES 입력 채널 수가 wafer-level etch spatial profile 예측 성능에 미치는 영향을 분석하였다.
% GPT2 small(6-layer)을 backbone으로 사용하고, patch 및 시계열 설정은 $p{=}48/s{=}24$, $N_T{=}6000$으로 고정한 상태에서, band selection 단계의 NMS window를 조절하여 OES 채널 수 $N_{\mathrm{oes}}$을 단계적으로 증가시켰다.
% Table~(\ref{tab:channel_sweep})에 나타낸 것과 같이 OES 채널 수 증가에 따라 shape 예측 성능은 전반적으로 개선되었으며, $N_{\mathrm{oes}}=72$에서 최저 shape MSE가 관측되었다.
% 반면 mean 예측 성능과 최종 etch depth MAE는 $N_{\mathrm{oes}}=48$에서 최적을 보였고, 채널 수를 72로 확장할 경우 소폭 저하되는 경향이 나타났다.
% 이는 다수의 파장이 공간 분포 형태 복원에는 기여하지만, mean과 같은 전역적 스칼라 특성 예측에서는 중복 정보 또는 노이즈로 작용할 수 있음을 시사한다.
This experiment examines the effect of the number of OES input channels on wafer-level etch spatial profile prediction.
Using GPT2-small (6 layers) as the backbone, we fix the patch and temporal settings to $L_p{=}48$, $S{=}24$, and $N_T{=}6000$, and progressively increase the number of OES channels $N_{\mathrm{oes}}$ by adjusting the NMS window in the channel selection stage.
As reported in Table~\ref{tab:channel_sweep}, spatial shape prediction accuracy improves overall as $N_{\mathrm{oes}}$ increases, achieving the lowest shape MSE at $N_{\mathrm{oes}}{=}72$.
In contrast, mean prediction accuracy and the final etch depth MAE are optimized at $N_{\mathrm{oes}}{=}48$ and exhibit a slight degradation when the channel count is further increased to 72.
These results suggest that while a larger set of wavelengths enhances reconstruction of spatial distribution patterns, excessive spectral information may introduce redundancy or noise for predicting global scalar quantities such as the wafer-level mean.

% Requires: \usepackage{booktabs}
\begin{table}[t]
    \centering
    \caption{10-fold cross-validation errors (mean $\pm$ std) for different numbers of OES channels.}
    \label{tab:channel_sweep}
    \footnotesize
    \setlength{\tabcolsep}{4pt}
    \renewcommand{\arraystretch}{1.12}
    \begin{tabular}{ccccc}
        \toprule
        $N_{\mathrm{oes}}$ & $N_{\mathrm{c}}$
        & MSE (shape)
        & MSE (mean)
        & MAE (etch) \\
        \midrule
        12 & 37 & $1.11 \pm 0.98$ & $2.60 \pm 3.76$ & $1.31 \pm 0.77$ \\
        24 & 49 & $0.42 \pm 0.22$ & $3.80 \pm 3.62$ & $1.52 \pm 0.73$ \\
        48 & 73 & $0.37 \pm 0.18$ & $\mathbf{2.28 \pm 2.00}$ & $\mathbf{1.25 \pm 0.51}$ \\
        72 & 97 & $\mathbf{0.23 \pm 0.08}$ & $2.69 \pm 2.31$ & $1.36 \pm 0.56$ \\
        \bottomrule
    \end{tabular}
\end{table}

\subsection{Backbone and Model Capacity Analysis}

\begin{table}[t]
    \centering
    \caption{10-fold cross-validation errors (mean $\pm$ std) for different model backbones and input channel configurations.}
    \label{tab:model_sweep_with_baseline}
    \footnotesize
    \setlength{\tabcolsep}{4pt}
    \renewcommand{\arraystretch}{1.15}
    \begin{tabular}{lccccc}
        \toprule
        \textbf{Backbone} 
        & \textbf{L}
        & \textbf{Ch}
        & \textbf{MSE (shape)} 
        & \textbf{MSE (mean)} 
        & \textbf{MAE (etch)} \\
        \midrule
        LLaMA & 12 & 49 & $\mathbf{0.19 \pm 0.07}$ & $3.20 \pm 2.58$ & $1.42 \pm 0.56$ \\
        LLaMA & 6  & 49 & $0.22 \pm 0.07$ & $\mathbf{2.89 \pm 2.14}$ & $\mathbf{1.41 \pm 0.49}$ \\
        LLaMA & 6  & 73 & $0.26 \pm 0.14$ & $3.23 \pm 2.87$ & $1.50 \pm 0.62$ \\
        GPT2  & 12 & 73 & $0.33 \pm 0.13$ & $3.59 \pm 1.92$ & $1.53 \pm 0.43$ \\
        GPT2  & 6  & 73 & $0.53 \pm 0.40$ & $3.70 \pm 3.29$ & $1.53 \pm 0.71$ \\
        \midrule
        \multicolumn{3}{l}{\textit{Global Mean Baseline}} 
        & $12.91 \pm 0.37$ 
        & $0.16 \pm 0.07$ 
        & $2.97 \pm 0.04$ \\
        \bottomrule
    \end{tabular}
\end{table}

% 본 실험에서는 backbone LLM의 종류, 모델 깊이, 그리고 입력 채널 수가 wafer-level etch spatial profile 예측 성능에 미치는 영향을 분석하였다.
% GPU 메모리 제약으로 context length가 약 300 token으로 제한된 조건에서, $N_T{=}6000$, $p{=}64/s{=}32$ 설정을 고정하여 backbone 및 모델 용량의 영향을 비교하였다.
% Table~\ref{tab:model_sweep_with_baseline}에서 보듯이, LLaMA 기반 모델은 GPT2 대비 shape MSE, mean MSE, etch MAE 전반에서 일관되게 우수한 성능을 보여, wafer-level 공간 분포 예측에서 backbone의 사전학습 표현력이 중요함을 확인하였다.
% 모델 깊이 측면에서는 shape 예측은 12-layer에서 유리한 반면, mean 및 MAE는 6-layer에서 더 안정적인 성능을 보였으며, 이는 공간 분포 형태와 전역 평균 예측이 서로 다른 표현 수준을 요구함을 시사한다.
% 입력 채널 수를 49에서 73으로 확장한 경우에도 성능 개선은 관찰되지 않았다.
% 한편 Global Mean Baseline은 mean MSE에서는 낮은 오차를 보이지만, wafer 내 공간 분포를 전혀 모델링하지 않으며, 제안한 LLaMA 기반 모델은 공간 분포를 명시적으로 예측하는 더 어려운 문제를 다루면서도 최종 etch depth MAE 기준에서 해당 baseline과 경쟁력 있는 성능을 달성하였다.
% 이는 본 모델이 단순 평균 추정을 넘어, in-situ 공정 시계열로부터 wafer-level 공간 정보를 실질적으로 복원할 수 있음을 보여준다.
This experiment evaluates the effects of backbone LLM choice, model depth, and input channel dimensionality on wafer-level etch spatial profile prediction.
Owing to GPU memory constraints, the context length is limited to approximately 300 tokens; accordingly, the input configuration is fixed to $N_T{=}6000$ with $L_p{=}64$ and $S{=}32$ to focus on backbone and capacity effects.
As reported in Table~\ref{tab:model_sweep_with_baseline}, LLaMA-based models consistently outperform GPT2 across shape MSE, mean MSE, and etch depth MAE, highlighting the critical role of pretrained backbone representations in spatial profile prediction.
Increasing model depth improves shape prediction, with LLaMA 12-layer models achieving lower shape error, whereas mean prediction and MAE are more stable under the LLaMA 6-layer setting, indicating that spatial structure and global magnitude rely on different representational capacities.
Increasing the number of input channels from 49 to 73 does not lead to further performance gains under these settings.
Although the Global Mean Baseline yields low mean MSE, it does not model spatial variation.
In contrast, the proposed LLaMA-based model explicitly predicts spatial profiles and achieves competitive etch depth MAE despite solving a more demanding task, demonstrating its ability to recover wafer-level spatial information from in-situ process time series beyond simple averaging.

% (optional)- $d_m, D, N_p(L_p, S)$ 변화에 따른 MAE 실험

% - 고전 신호처리/ML 알고리즘 구현과 성능 비교

% - 관련 연구와의 성능 비교
% \textcolor{red}{
  % * 데이터셋 자체 공개가 안되어 있음. BOSCH DRIE 데이터셋은 공개(Sep 15, 2025)된지 6개월이 채 되지 않아, 이를 활용한 발표된 연구가 없음.
  % (To the best of our knowledge, this is the first study to utilize the BOSCH plasma-etching dataset for spatial profile prediction. - introduction으로)}

%\section{Application and Discussion}
%- conclusion 쪽에서 언급. application 일부는 dateset의 target domain에서 언급해도 될 듯
%Time-LLM-Based Prediction of Wafer-Level Etch Quality from In-Situ Process Time-Series %Signals 기술을 어떻게 활용할 수 있는지 그리고 관련 discussion 내용

\section{Conclusion}

%본 논문에서는 대규모 언어 모델의 표현력을 반도체 공정 감시에 확장하기 위해 Time-LLM 기반 공간 회귀 모델을 제안하였다. 
%In-situ 공정 파라미터(PP)와 광학 방출 분광(OES) 신호로부터 웨이퍼 내 89개 지점의 식각 깊이 분포를 직접 예측하도록 설계하였으며, 전역 평균(mean)과 공간 분포 형태(shape)를 분리 추정함으로써 공정 드리프트와 공간적 비균일성을 동시에 정밀하게 포착할 수 있도록 하였다.

%우리가 아는 한, 본 연구는 실제 공정에서 수집된 BOSCH 플라즈마 식각 데이터셋을 활용하여 Time-LLM 기반 시계열 재프로그래밍 접근을 wafer-level 공간 프로파일 예측 문제에 체계적으로 적용하고 실증적으로 검증한 초기 사례에 해당한다. 
%실험 결과, 제안 모델은 데이터가 제한된 환경에서도 안정적이고 일관된 성능을 보였으며, 복합적인 2차원 공간 변동을 효과적으로 모델링하였다. 이는 LLM 기반 시계열 모델이 단순 예측을 넘어 공간적 품질 특성까지 직접 추정할 수 있음을 보여주며, 향후 In-situ 공정 모니터링 시스템과 결합될 경우 실시간 수율 관리 및 조기 이상 감지를 위한 지능형 감시 프레임워크로 확장될 수 있는 가능성을 제시한다. 향후 연구에서는 다양한 공정 조건과 확장된 데이터 환경에서의 일반화 성능 및 실시간 추론 효율성을 추가적으로 분석할 예정이다.

% In this paper, we proposed a Time-LLM–based spatial regression model that predicts wafer-level etch depth distributions at 89 spatial locations directly from in-situ process parameters and OES time-series data. 
% By decomposing the prediction into global mean and spatial shape components, the model effectively captures both process-level drift and intra-wafer non-uniformity. 
% Experiments on the BOSCH plasma-etching dataset demonstrate accurate and robust prediction of two-dimensional wafer-level spatial patterns under data-limited conditions, highlighting the potential of LLM-based reprogramming for wafer-level spatial modeling.

In this paper, we proposed a Time-LLM–based spatial regression model for predicting wafer-level etch depth distributions at 89 spatial locations directly from in-situ process parameter and OES time-series data.
By decomposing the prediction into global mean and spatial shape components, the proposed model jointly captures process-level drift and intra-wafer spatial non-uniformity in a structured and interpretable manner.
To the best of our knowledge, this study represents one of the first empirical investigations that leverage the BOSCH plasma-etching dataset for wafer-level spatial profile prediction using an LLM-based reprogramming framework.
Experimental results demonstrate that the proposed model can stably recover two-dimensional wafer-level etch patterns under data-limited conditions, while achieving competitive accuracy in wafer-level prediction.
These findings provide quantitative evidence that LLM-based reprogramming constitutes a viable and effective approach for modeling wafer-level spatial characteristics in plasma etching processes, extending beyond conventional scalar virtual metrology paradigms.

\bibliographystyle{IEEEtran}
%\bibliography{main}
\bibliography{main-short}
%\bibliography{main-full-author}

\end{document}